\documentclass{aa}  
\usepackage{graphicx}
\usepackage{txfonts}
\usepackage{amsmath}
\usepackage{amssymb}
\usepackage{hyperref}
\begin{document}

   \title{The elusive synchrotron precursor of collisionless shocks}

   \author{Ruben Zakine
          \inst{1,2}
          \and
          Martin Lemoine \inst{3}
          }

   \institute{\'Ecole polytechnique, F-91128 Palaiseau Cedex, France
     \and
     Institut d'Astrophysique de Paris,
     CNRS -- UPMC, 
     98 bis boulevard Arago, 75014 Paris, France\\
              \email{ruben.zakine@univ-paris-diderot.fr}
         \and
             Institut d'Astrophysique de Paris,
     CNRS -- UPMC, 
     98 bis boulevard Arago, 75014 Paris, France\\
             \email{lemoine@iap.fr}
             }

   \date{}

% \abstract{}{}{}{}{} 
% 5 {} token are mandatory
 
  \abstract
  {Most of the plasma microphysics which shapes the acceleration
    process of particles at collisionless shock waves takes place in
    the cosmic-ray precursor, upstream of the shock front, through the
    interaction of accelerated particles with the unshocked plasma.}
  {Detecting directly or indirectly the
  synchrotron radiation of accelerated particles in this synchrotron
  precursor would open a new window on the microphysics of
  acceleration and of collisionless shock waves. Whether such a
  detection is feasible is discussed in the present paper.}
 {To this effect, we provide analytical estimates of the spectrum and
   of the polarization fraction of the synchrotron precursor for both
   relativistic and non-relativistic collisionless shock fronts,
   accounting for the self-generation or amplification of magnetic
   turbulence.}
 {In relativistic sources, the spectrum of the precursor is harder
   than that of the shocked plasma because the upstream residence
   time increases with particle energy, leading to an effectively hard
   spectrum of accelerated particles in the precursor volume. At high
   frequencies, typically in the optical to X-ray range, the
   contribution of the precursor becomes sizeable,  but we find that in
   most cases studied, it remains dominated by the synchrotron or
   inverse Compton contribution of the shocked plasma; its
   contribution might be detectable only in trans-relativistic shock
   waves. Non-relativistic sources offer the possibility of spectral
   imaging of the precursor by viewing the shock front edge-on.  We
   calculate this spectro-morphological contribution for various
   parameters. The synchrotron contribution is also sizeable at the
   highest frequencies (X-ray range), corresponding to maximum energy
   electrons propagating on distance scales $\sim 10^{16}\,$cm away
   from the shock front. If the turbulence is tangled in the plane
   transverse to the shock front, the resulting synchrotron radiation
   should be nearly maximally linearly polarized; polarimetry thus
   arises as an interesting tool to reveal this precursor.}
   {}


   \maketitle
%
%________________________________________________________________

\section{Introduction}\label{sec:introd}
Most of the non-thermal emission seen in powerful astrophysical
objects such as supernovae remnants (SNR), pulsar wind nebulae, and
gamma-ray bursts is believed to arise through the synchrotron and
inverse Compton radiation of electrons accelerated at a strong
collisionless shock
front~\citep[e.g.][]{1987PhR...154....1B,2012SSRv..173..309B}.

It has been long recognized that the acceleration of particles is
intimately related to the self-generation of turbulence or to the
amplification of pre-existing magnetic fluctuations in the shock
precursor. This precursor delimits the region of space upstream of the
shock where the particles that are accelerated at the shock front
interact with the background plasma. In this context, the observation
of radiative signatures from this precursor would bring new and
unprecedented information on the turbulence, on the acceleration
process and the shock physics, which is currently probed indirectly
through the radiation of the shocked plasma on much greater space and
time scales.  Among the top questions to be addressed are the
following: What is the magnetization of the background flow? What is
the level of turbulence and on what scales is it distributed? How are
the accelerated particles transported in the self-generated/amplified
turbulence?

The precursor is bounded by the length over which the accelerated
particles return to the shock front after scattering in the ambient
magnetized flow. The hierarchy between the typical scale of this
precursor ($\ell_{\rm p}$) and that of the shocked downstream region
($\ell_{\rm d}\,\sim\,R$, $R$ radius of the shock) -- namely
$\ell_{\rm p}\,\ll\,\ell_{\rm d}$ -- is precisely what makes the
detection of the precursor difficult. Its synchrotron contribution has
been debated in the literature. For instance,
\citet{2009ApJ...707L..92S} observe in their numerical simulation of a
relativistic shock front a net signal from the shock precursor, at
higher energies than the radiation from the downstream flow, while
\citet{2008PhRvE..77b6403G} have argued that the strong enhancement of
the turbulence in the transition layer of a relativistic collisionless
shock should leave a distinct imprint in the synchrotron spectrum. In
sub-relativistic shock waves, rather broad shock transitions seen in
X-ray have been interpreted as the signature of an extended
synchrotron precursor, although this broadening might otherwise
represent deviations of the shock front away from sphericity
\citep{2013ASPC..470..269V}. Numerical simulations of such X-ray
images, including a contribution from the synchrotron precursor
indicate that the possibility of detection is directly related to the
unknown length scale of the precursor, and the feasibility of a
detection remains uncertain (see
e.g. \citealt{1994A&A...281..220A,1994AstL...20..157Az,1998ApJ...493..375R,2003A&A...412L..11B,2003ApJ...586.1162L,2005ApJ...632..920E,2010MNRAS.405L..21M}).

The main objectives of this paper are to characterize the synchrotron
signature of the precursor and to assert the feasibility of a
detection, in both relativistic and non-relativistic sources. We also
discuss the degree of polarization, since it may prove an essential
tool in extracting the precursor signal. In the relativistic regime,
a Lorentz aberration generically implies a configuration in which the
source is not spatially resolved and where the flow and the line of
sight are nearly coincident. Consequently, the synchrotron
contribution must be extracted from the overall spectral energy
distribution. As we show in Sect.~\ref{sec:rel}, the precursor spectrum
is generally much harder than that of the shocked region, suggesting
that a detection, if feasible, should occur at the highest
frequencies, typically in the X-ray range.

In the non-relativistic regime, the shock front can be seen edge-on,
which offers the possibility of direct imaging of the precursor. We
calculate this spectro-morphological contribution in detail in
Sec.~\ref{sec:nrel}; in particular, we argue that the detection of
linear polarization on small angular scales upstream of the shock
would provide a net signature of this precursor. We summarize our
results and provide conclusions in Sec.~\ref{sec:conc}.

\section{Precursor of relativistic shock fronts}\label{sec:rel}
We consider a relativistic source, with the observer on-axis, as is the
case of a gamma-ray burst afterglow. On both analytical and numerical
grounds, little is known about the structure of the precursor on the
space and time scales that are representative of astrophysical
objects. We thus consider a general picture, in which both the
precursor and the shocked downstream plasma are populated by a
background magnetic field and by a magnetized turbulence.  The
fraction of local total energy density stored in magnetic turbulence
is written $\epsilon_{B{\rm d}}$ downstream and $\epsilon_{B{\rm p}}$
in the precursor, while $\sigma$ represents the magnetization of the
background flow. Consequently, the comoving downstream magnetic field reads
$\delta B_{\rm\vert d}\,=\,\left(\epsilon_{B{\rm d}}8\pi\,2\Gamma_{\rm
  sh}^2\,n_{\rm u} m_p c^2\right)^{1/2}$, while the precursor frame
turbulent field reads $\delta B_{\rm\vert p}\,=\,\left(\epsilon_{B{\rm
    p}}8\pi\,n_{\rm u} m_p c^2\right)^{1/2}$ (see below for the
association of the precursor frame with the upstream frame) and the
background upstream frame field $B_{\rm\vert u}\,=\,\left(\sigma
8\pi\,n_{\rm u} m_p c^2\right)^{1/2}$; $n_{\rm u}$ denotes the proper
density of the unshocked plasma and $\Gamma_{\rm sh}$ the bulk Lorentz
factor of the shock front relative to the background plasma.

For simplicity, and to reduce the number of free parameters, we
consider here a turbulence of homogeneous power both upstream and
downstream. Both theory and phenomenology suggest that the downstream
turbulence should decay away from the shock front through
collisionless damping (see
e.g. \citealt{2008ApJ...674..378C,2013MNRAS.428..845L,2013MNRAS.435.3009L,2015JPlPh..81a4501L,2015MNRAS.453.3772L}). However,
at late observing times, such as those considered here, the resulting
synchrotron self-Compton spectrum can be approximated with an
effective $\epsilon_{B{\rm d}}$ parameter of low value, representative
in some way of the average fraction over the blast length scale (see
\citealt{2015MNRAS.453.3772L}). Considering the distribution of
magnetic energy in the shock precursor, we take into account its
possible evolution with distance through the average value of
$\epsilon_{B{\rm p}}$ and through two extreme models, which are
discussed below.

An important issue in the calculation of the synchrotron emission and
its polarization is the choice of reference frame. In the background
magnetic field, the relevant frame for the synchrotron process is the
upstream frame (neglecting any deceleration of the flow on the scale
of the precursor). In weakly magnetized shock waves, the turbulence is
believed to be self-generated through filamentation-type instabilities
(e.g. \citealt{1999ApJ...526..697M}). For the particular case of the
Weibel instability, there is a frame in which the turbulence is mostly
magnetic in nature; it can be shown that this frame moves at
sub-relativistic velocities with respect to the background
plasma~\citep{Pea2016}. Up to this sub-relativistic motion the
relevant frame is again the upstream frame, and in this frame the
accelerated electron population is strongly beamed forward into an
angle $\sim\, 1/\Gamma_{\rm sh}$.

The accelerated population is modelled with a power law between a
minimum Lorentz factor $\gamma_{\rm m\vert p}\,\sim\,2\Gamma_{\rm sh}
\gamma_{\rm m}$ ($\gamma_{\rm m}$ being the corresponding shock frame
quantity), a spectral index $s$, and a maximum  Lorentz factor
determined by energy losses. The minimum Lorentz factor of the
accelerated population is defined as usual
by~\citep[e.g.][]{2004RvMP...76.1143P} $\gamma_{\rm m}\,\simeq\,\vert
s-2\vert/\vert s-1\vert\,\,\epsilon_e \Gamma_{\rm sh}m_p/m_e$ so that
the accelerated particle population is described, in the shock frame,
by
\begin{equation}
  \frac{{\rm d}n_{e\vert\rm sh}}{{\rm d}\gamma_e}\,=\,\frac{\epsilon_e
    e_{\rm sh}}{\gamma_{\rm m} m_e c^2}\frac{\vert
    s-2\vert}{\gamma_{\rm m}}\left(\frac{\gamma_e}{\gamma_{\rm
      m}}\right)^{-s}
\end{equation}
This guarantees that the total energy density of the accelerated
electron population is normalized to $\epsilon_e e_{\rm sh}$, with
$e_{\rm sh}$ the total energy density at the shock, while the number
density of electrons is normalized to $e_{\rm sh}/\left(\Gamma_{\rm
  sh}m_p c^2\right)$, as expected from the shock crossing conditions.

The electron radiation in the micro-turbulence can be described by a
synchrotron process because the deflection of the pitch angle by an
angle $\delta\vartheta\,=\,e\delta B\lambda_{\delta B}/\left(\gamma_e
m_ec^2\right)$ over a coherence length $\lambda_{\delta B}$ is
much larger than the angle $1/\gamma_e$ that determines the opening
angle of the emission cone of radiation, see
e.g. \citet{2013ApJ...774...61K}.  Indeed,
$\gamma_e\delta\vartheta\,\sim\, 80 \epsilon_{B\rm
  p,-5}^{1/2}\lambda_{\delta B,1}\gamma_e/\gamma_{\rm m\vert p}$, with
$\lambda_{\delta B,1}\,=\,0.1\lambda_{\delta B}c/\omega_{\rm pi}$ the
coherence length scale in units of tens of skin depth scales; unless
otherwise noted, we use cgs units with $Q_x\,\equiv\,Q/10^x$ for any
quantity $Q$.

Owing to Lorentz beaming of the electron population, we only consider  the
radiation emitted by the particles contained within an angle
$\Delta\theta\,=\,1/\Gamma_{\rm sh}$ away from the line of sight, as
measured from the source position. The synchrotron spectral power per
unit solid angle of the precursor can then be expressed as
\begin{equation}
  \frac{{\rm d}P}{{\rm d}\nu{\rm d}\Omega}\,=\,\int{\rm d}r r^2\,{\rm
    d}\gamma_e\,\, \frac{{\rm d}n_{e\vert\rm p}}{{\rm d}\gamma_e}\,p_\nu
  \label{eq:sp}
\end{equation}
in terms of ${\rm d}n_{e\vert\rm p}/{\rm d}\gamma_e$ the electron
distribution function in the precursor frame, and $p_\nu$ the
isotropic spectral power of one
electron~\citep{1965ARA&A...3..297G,1968ApJ...154..499L}:
$p_\nu\,=\,\sqrt{3}e^3 B_\perp/(m_e c^2)F(z)$ with $F(z)\,\equiv\,z
\int_z^{+\infty}{\rm d}\eta\, K_{5/3}(\eta)$, and
$z\,\equiv\,\nu/\nu_{\rm c}$, $\nu_{\rm c}\,\equiv\,
3eB_\perp\gamma_e^2/(4\pi m_e c)$ the peak frequency of synchrotron
emission for an electron of Lorentz factor $\gamma_e$.

Solid angles do not appear in Eq.~(\ref{eq:sp}) because their product
cancels out to unity; this product involves the solid angle of the
volume element of the precursor
($\Delta\Omega\,\simeq\,\pi\Delta\theta^2$), the probability that an
electron inside the precursor radiates towards the observer
($p_\rightarrow\,\simeq\,\delta\Omega/\Delta\Omega$) and
$1/\delta\Omega\,\simeq\,\gamma_e^2/\pi$, which is the inverse solid
angle of the emission of one electron along its near ballistic
trajectory.

The distribution function of the electrons inside the precursor can be
formally determined through a Lorentz transform of the corresponding
distribution function in the shock frame, and the latter can be
obtained by solving a stationary transport equation with a loss term
corresponding to the finite residence time spent in the
precursor. Here, we approximate this distribution function in the
precursor frame as follows:
\begin{equation}
  \frac{{\rm d}n_{e\vert\rm p}}{{\rm
      d}\gamma_e}\,\simeq\,\frac{R^2}{r^2}\, \frac{\vert
    s-1\vert}{\gamma_{\rm m\vert p}}\left(\frac{\gamma_e}{\gamma_{\rm
      m\vert p}}\right)^{-s}\, 2\Gamma_{\rm sh}^2n_{\rm
    u}\,\Theta\left[r_{\rm p}(\gamma_e)-r\right]
  \label{eq:dndg1}
\end{equation}
The Heaviside function models the finite length scale up to which
particles of a given Lorentz factor can travel. This precursor length
scale is related to the (upstream frame) residence time $t_{\rm res}$
through $r_{\rm p}\,=\,R+c t_{\rm res}\left(1-\beta_{\rm
  sh}\right)\,\simeq\,R+ct_{\rm res}/(2\Gamma_{\rm sh}^2)$.  The
residence time can be calculated as the time it takes to deflect the
accelerated electron by an angle $\sim\,1/\Gamma_{\rm
  sh}$~\citep{2001MNRAS.328..393A,2006ApJ...651..979M,2009MNRAS.393..587P}. If
the background magnetic field controls the transport of the
accelerated electrons, this residence time is $t_{\rm
  res}^{(1)}\,\simeq\,\Gamma_{\rm sh}^{-1}\gamma_e m_e c/(e B_{\rm
  u})$. In contrast, if pitch angle scattering in a micro-turbulence
of length scale $\lambda_{\delta B}$ governs the return of particles
to the shock, $t_{\rm res}^{(2)}\,\simeq\,\Gamma_{\rm
  sh}^{-2}\gamma_e^2m_e^2c^3/(\lambda_{\delta B}e^2\delta B_{\rm
  p}^2)$. Whether one or the other occurs depends on $\gamma_e$ and
the hierarchy between $B_{\rm u}$ and $\delta B_{\rm p}$. As the
scattering frequency in a micro-turbulence scales as $\gamma_e^{-2}$,
while the gyrofrequency scales with $\gamma_e^{-1}$, we expect the
background field to control the residence time at higher energies. In
order to bracket this realistic scenario, we consider in the following
the above two extremes: (1) where regular deflection in a background
field dominates at all energies and (2) where stochastic deflection in
a small scale micro-turbulence of uniform energy density dominates at
all energies.  Correspondingly, we write $r_{\rm
  p}(\gamma_e)=R+\Delta_i\gamma_e^i$, with $i=1,2$ for models (1) or
(2); $\Delta_i=ct_{\rm res}^{(i)}/(2\Gamma_{\rm sh}^2)$.

In principle, to perform an accurate calculation of the synchrotron
spectrum, one would need to specify both the residence time and the
scaling of the micro-turbulent field with distance away from the
shock. In order to bracket the uncertainty associated with transport and
the evolution of the magnetic energy in the precursor, we use both
extreme models (1) and (2). At small frequencies, corresponding to
particles of small Lorentz factor exploring the near precursor, the
transport should be governed by the stochastic field so that the
synchrotron flux should be adequately described by model $(2)$. The
high-frequency part, however, corresponds to  high-energy particles
spending most of their time in the far precursor, where presumably the
background magnetic field dominates the transport and the radiation
process. One should thus expect the high-frequency part of the
synchrotron spectrum to be described by model $(1)$. These two extreme
models should thus bracket adequately the synchrotron spectrum of the
precursor.

Because the synchrotron flux scales with residence time, and because
particles of larger Lorentz factor spend a greater amount of time in
the precursor, the spectral power is expected to be harder than the
standard slow-cooling spectrum of a population trapped in the
downstream. This spectral dependence indeed derives from the
integration of Eq.~(\ref{eq:sp}), giving a spectral flux on the
detector
\begin{equation}
  F_{\nu,\rm p}\,\simeq\,\frac{\sqrt{3}\vert s-1\vert(1+z)}{d_{\rm L}^2}R^2\Delta_i n_{\rm
    u}\frac{e^3 B_\perp}{m_e c^2} \,\int{\rm d}\gamma_e\,\gamma_{\rm m\vert p}^{-1}
  \left(\frac{\gamma_e}{\gamma_{\rm
      m\vert p}}\right)^{i-s}F(z)
\label{eq:sp2}
,\end{equation}
which leads to a hard particle spectrum as integrated over the
precursor length scale. Henceforth, all frequencies are written in the
observer frame and $\Delta_i$ is evaluated at $\gamma_{\rm m\vert p}$.

If $i=1$, the last integral over $\gamma$ can be performed using the
standard results of \citet{1965ARA&A...3..297G} and
\citet{1968ApJ...154..499L}, leading to a spectrum $F_{\nu,\rm
  p}\,\propto\,\nu^{(2-s)/2}$ in the inertial range -- between the
synchrotron peak frequencies $\nu_{\rm m\vert p}$ and $\nu^{(1)}_{\rm
  max\vert p}$ -- of particles of respective Lorentz factor
$\gamma_{\rm m\vert p}$ and $\gamma_{\rm max}$; of course, $F_{\nu,\rm
  p}\,\propto\,\nu^{1/3}$ below $\nu_{\rm m\vert p}$. Hence, for
$s\,\simeq\,2.3$, $\nu F_{\nu,\rm p}\,\propto\, \nu^{0.85}$ up to the
highest frequencies. If $i=2$, these results are not applicable unless
$s\,>\,7/3$; if $s\,<\,7/3$, the spectrum at any frequency $\nu$ is
indeed dominated by the $1/3$ low-frequency extension of the spectrum
of particles radiating with a synchrotron peak frequency larger than
$\nu$. To calculate this contribution, the function $F(z)$ can be
approximated by its small $z$ limit,
$F(z)\,\simeq\,2^{-1/3}3\Gamma(5/3)\,z^{1/3}$ ($z\,\ll\,1$) and the
integration can be performed over the effective distribution of
particles in the precursor. 

This results in a broken power-law approximation to the spectral flux
of the precursor
\begin{eqnarray}
   F_{\nu,\rm p}&\,\simeq\,&\frac{\varphi_i(1+z)}{d_{\rm L}^2}R^2c t_{\rm res}^{(i)}(\gamma_{\rm m\vert p})\,
   n_{\rm u}\frac{e^3 B_\perp}{m_e c^2}\,\,
   \Theta\left[\nu^{(i)}_{\rm max\vert p}-\nu\right]\,\nonumber\\
&&\times\left\{\Theta\left[\nu-\nu_{\rm m\vert
         p}\right]\left(\frac{\nu}{\nu_{\rm m\vert p}}\right)^{a_i}+
       \Theta\left[\nu_{\rm m\vert p}-
         \nu\right]\left(\frac{\nu}{\nu_{\rm m\vert p}}\right)^{1/3}
    \right\}
  \label{eq:sp3}
,\end{eqnarray}
where, as before, $i\,=\,1,\,\,2$ and
\begin{eqnarray}
  \varphi_1&\,=\,&\frac{\vert s-1\vert\sqrt{3}}{4}2^{(s-2)/2}\frac{4+3s}{3s}
  \Gamma\left[\frac{3s-4}{12}\right]
  \Gamma\left[\frac{3s+4}{12}\right]
  \nonumber\\
  a_1&\,=\,&\frac{2-s}{2}\nonumber\\
  \varphi_2&\,=\,&\frac{3.7\vert s-1\vert}{7/3-s}\left[
  \left(\frac{\gamma_{\rm max\vert p}}{\gamma_{\rm m\vert p}}\right)^{7/3-s}-1\right]\nonumber\\
  a_2&\,=\,& \frac{1}{3}
  \label{eq:norm}
\end{eqnarray}
The last two expressions assume $s\,<\,7/3$.  We note that the
numerical prefactor $\varphi_2$ can become large if $s$ comes close to
$7/3$, where the contributions of all particles add up to form the
synchrotron spectrum at low frequencies. For $s\,=\,7/3$, the factor
$\vert s-7/3\vert^{-1} \left[\left(\gamma_{\rm max\vert p}/\gamma_{\rm
    m\vert
    p}\right)^{s-7/3}-1\right]\,\rightarrow\,\ln\left(\gamma_{\rm
  max\vert p}/\gamma_{\rm m\vert p}\right)$.

It is possible to  compare the precursor synchrotron flux to that produced
downstream. Following \citet{2000ApJ...543...66P}, the latter can be
expressed as $F_{\nu,\rm d}\,\simeq\,0.26(1+z)\,d_{\rm L}^{-2}\Gamma_{\rm
  sh}n_{\rm u}R^3e^3B_{\rm d}/(m_e c^2)\left(\nu/\nu_{\rm
  m}\right)^{(1-s)/2}$ for $\nu\,>\,\nu_{\rm m}$ (assuming the
slow-cooling regime and a homogeneous circumburst medium, $\nu_{\rm
  m}$ representing the downstream minimum synchrotron
frequency); $B_{\rm d}$ being a factor $\Gamma_{\rm sh}$ larger than
$B_{\rm u}$ (both measured in their own comoving frames), $\nu_{\rm
  m\vert p}\,\simeq\,\nu_{\rm m}\left(\epsilon_{B{\rm
    p}}/\epsilon_{B{\rm d}}\right)^{1/2}$; consequently, the upstream
contribution appears suppressed by a factor $\Delta_i/R$ at the
minimum frequency (at equal $\epsilon_{B\rm p}$ and $\epsilon_{B\rm
  d}$) up to the numerical prefactors, which can be traced back to the
difference of volume occupied by the radiating electrons. At high
frequencies,  however, a larger relative contribution is expected from
the precursor, due to the rising spectrum detailed above.

A detailed comparison thus requires a careful determination of the
maximum Lorentz factor up to which particles can be accelerated, as
well as an evaluation of the inverse Compton contribution from the
downstream plasma, which takes over the synchrotron radiation at high
frequencies.

The maximum Lorentz factor is determined by the competition between
the residence time (both upstream and downstream) and the synchrotron
or inverse Compton loss time. Given that the energy gain is of order
unity per Fermi cycle, the acceleration time scale corresponds to the
greater of the downstream or the upstream residence time (as evaluated
in the same frame). Below, we instead calculate the maximum Lorentz
factor associated with transport either downstream or upstream and
keep the smaller of the two values. In a generic setting, inverse Compton
losses dominate both downstream and upstream
\citep{2006ApJ...651..328L}.

We consider first the maximum energy for downstream particles, calculated
in the downstream comoving frame. Assuming that particles scatter in a
micro-turbulence of coherence length $\lambda_{\delta B}$, the maximum
Lorentz factor for electrons corresponding to synchrotron and inverse
Compton losses is
\begin{equation}
  \gamma_{\rm max\vert d}\,\simeq\,\left[\frac{9}{4}\frac{\lambda_{\delta B}}{r_e}\frac{1}{1+Y}\right]^{1/3},\nonumber\\
  \label{eq:gmd}
\end{equation}
where $r_e\,\simeq\,2.8\times 10^{-13}\,$cm denotes the classical
electron radius. A numerical estimate is $\gamma_{e,\rm max\vert
  d}\,\simeq\, 10^7\left(1+Y\right)^{-1/3}n_0^{-1/6}$. The $Y$ Compton
parameter is determined through~\citep{2001ApJ...548..787S}
$Y\,\simeq\,\left(\epsilon_e/\epsilon_{B_{\rm
    d}}\right)^{1/2}\left(\gamma_{\rm c,syn}/\gamma_{\rm
  m}\right)^{1-s/2}$, with $\gamma_{\rm c,syn}$ the cooling Lorentz
factor determined by synchrotron losses alone, i.e. the electron
Lorentz factor (in the downstream frame) for which the synchrotron
cooling length becomes comparable to the dynamical time scale $t_{\rm
  dyn}\,\simeq\,R/(\Gamma_{\rm sh}c)$. We ignore Klein-Nishina
suppression effects in the determination of $\gamma_{e,\rm max\vert
  d}$; they may be important at very early observer times, leading to
an effective $Y\,\sim\,\mathcal{O}(1)$, but become moderate at late
times~\citep{2010ApJ...712.1232W,2013MNRAS.435.3009L,2015MNRAS.453.3772L}.
 
As discussed in various studies (see
\citealt{2009MNRAS.393..587P,2012SSRv..173..309B} and references therein)
the above provides the correct way of defining the maximum Lorentz
factor for acceleration. For reference, however, we also define a
maximum Lorentz factor corresponding to the Bohm assumption, according
to which the scattering frequency is given by the gyrofrequency:
\begin{equation}
  \gamma^{\rm Bohm}_{\rm max\vert d}\,\simeq\, \left[\frac{6\pi e}{\sigma_{\rm T}\delta B_{\rm d}\left(1+Y\right)}\right]^{1/2}
  \label{eq:gmdB}
\end{equation}

We consider now the maximum Lorentz factor corresponding to the
upstream residence time, as evaluated in the upstream reference
frame. As discussed by \citet{2006ApJ...651..328L}, the energy loss is
dominated by inverse Compton radiation emitted by downstream electrons
and it is more convenient to calculate the maximum Lorentz factor in
the shock frame and then to transform back to the upstream frame. If
transport at the maximum energy is dominated by scattering in the
small-scale turbulence, corresponding to our model (2), the maximum
Lorentz factor is
\begin{equation}
  \gamma^{(2)}_{\rm max\vert p}\,\simeq\, \Gamma_{\rm
    sh}^{2/3}\left(\frac{\epsilon_{B{\rm u}}}{\epsilon_{B{\rm
        d}}}\right)^{1/3}\gamma_{e,\rm max\vert d}
  \label{eq:gmaxu1}
,\end{equation}
while in model (1), where the transport is governed by the orbit of
the accelerated particle in the background field,
\begin{equation}
  \gamma^{(1)}_{\rm max\vert p}\,\simeq\, \Gamma_{\rm
    sh}\left(\frac{\sigma}{\epsilon_{B{\rm
        d}}}\right)^{1/4}\gamma^{\rm Bohm}_{e,\rm max\vert d}
  \label{eq:gmaxu2}
\end{equation}
In principle, one must also bound these maximum Lorentz factors using
the condition that the upstream residence time cannot exceed the age
of the shock front in the upstream frame, i.e. $R/c$; for the range of
parameters that we consider here, however, this constraint is not
significant.

For reference, the dependence of these maximum Lorentz factors on the
shock parameters are (in the precursor frame)
\begin{eqnarray}
  2\Gamma_{\rm sh}\gamma_{\rm max\vert d}&\,\simeq\,& 2\times10^8\,E_{54}^{0.1}\lambda_{\delta B,1}^{1/3}\epsilon_{B{\rm d},-5}^{0.1}n_0^{-0.3}\epsilon_{e,-1}^{-0.2}t_{\rm obs,4}^{-0.4}z_+^{0.4}\nonumber\\
  \gamma^{(1)}_{\rm max\vert p}&\,\simeq\,& 4\times 10^8\,\sigma_{\rm u,-9}^{1/4}n_0^{-0.4}\epsilon_{B{\rm d},-5}^{-0.3}\epsilon_{e,-1}^{-0.3}t_{\rm obs,4}^{-0.2}z_+^{0.2}\nonumber\\
  \gamma^{(2)}_{\rm max\vert p}&\,\simeq\,&4\times 10^7 \epsilon_{B{\rm u}-5}^{1/3}\lambda_{\delta B,1}^{1/3}n_0^{-0.3}\epsilon_{B{\rm d}-5}^{-0.2}\epsilon_{e,-1}^{-0.2}t_{\rm obs,4}^{-0.2}z_+^{0.2}
  \label{eq:numgm}
\end{eqnarray}
with $z_+\,\equiv\,(1+z)/2$. This indicates that these Lorentz factors
are typically of the same order of magnitude. The above expressions
assume the limit $Y\,\gg\,1$ and the Lorentz factor of the blast is
related to the observer time $t_{\rm obs}$ and other standard
parameters following the standard \cite{1976PhFl...19.1130B}
deceleration law. We recall that in model $(i)$, it is necessary to retain the
minimum of $2\Gamma_{\rm sh}\gamma_{e,\rm max\vert d}$ and
$\gamma^{(i)}_{e,\rm max\vert p}$.

\begin{figure}[ht]
\includegraphics[width=0.95\columnwidth]{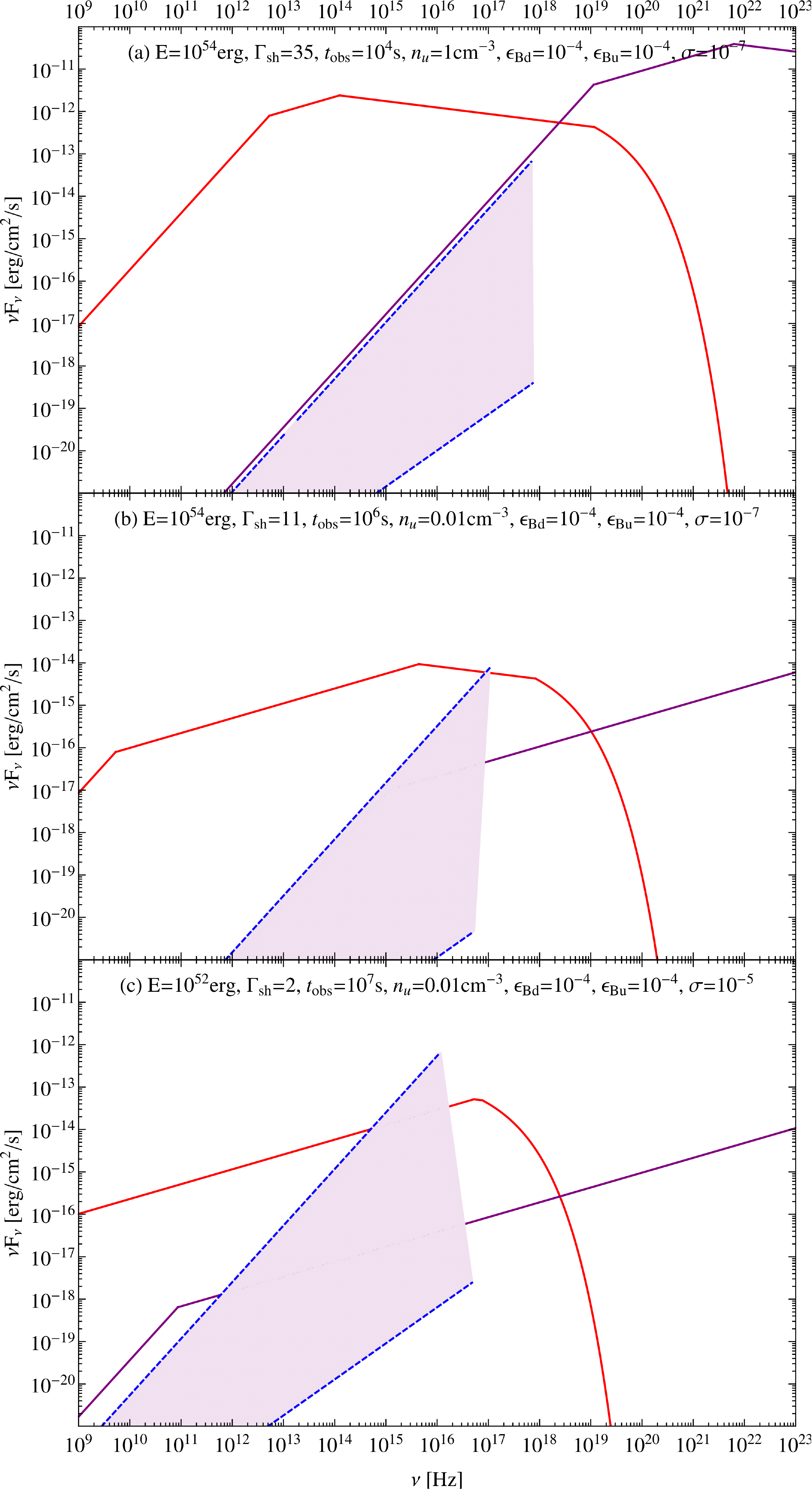}
  \caption{Spectral flux per log frequency interval $\nu F_\nu$ for
    the downstream synchrotron emission (red), the downstream inverse
    Compton spectrum (purple), and the precursor synchrotron signal (shaded
    area). This last shaded area is bracketed by the two extreme
    models of regular deflection [lower curve, model $(1)$] or
    stochastic small-angle deflection [upper curve, model $(2)$]. The
    parameters of the model are indicated in the figure. For panels
    (a) and (b), representative of the external shock of a long
    gamma-ray burst at different times, the redshift is $z=1$; for
    panel (c), representative of a low-luminosity gamma-ray burst or
    trans-relativistic supernova, $z=0.01$.
    \label{fig:relres}}
\end{figure}

Combining the above results, it is possible to plot the synchrotron self-Compton
(SSC) spectrum of the downstream radiation together with the
synchrotron flux from the precursor. The inverse Compton emission is
modelled with broken power laws following
\citet{2001ApJ...548..787S}. Examples are shown in
Fig.~\ref{fig:relres} for various cases of interest.

The synchrotron radiation from the precursor is indicated by the
shaded area, bracketed by the two models $(1)$ and $(2)$ corresponding
to regular deflection and radiation in a background magnetic field or
to stochastic small-angle deflection and radiation in a
micro-turbulent field; the lower curve corrresponds to $(1)$, while
the upper curve corresponds to $(2)$. Equations~(\ref{eq:numgm}) can
be used to scale the maximum Lorentz factor, hence frequency, to other
cases.

These results indicate that the precursor contribution is dominated by
the SSC spectrum of the shocked plasma in most cases considered.  It
appears that the only case in which this precursor can emerge is for a
trans-relativistic shock, as depicted in case (c), if the
micro-turbulent component dominates the scattering and the radiation
up to the maximum energies. In all cases the contribution of model
$(1)$ is strongly suppressed: although the spectrum is harder than
that of downstream, the maximum energy of particles remains limited by
the downstream residence time and the maximum frequency emitted
upstream is then suppressed by the relatively low magnetic field.

At extremely early times, the ratio $\Delta_i/R$ plays in favour of the
precursor contribution. This likely explains why the precursor
emission stands out clearly in the numerical simulations of
\citet{2009ApJ...707L..92S};  in such simulations the size of
the precursor is comparable to the size of the downstream shocked
plasma. We also note that at the early time of these simulations, the
accelerated particles concentrate in a layer in front of the shock,
i.e. the precursor has not developed to a full extent with a position
dependent minimum momentum; this explains the standard synchrotron
spectrum seen in such simulations. We have not considered such early
times in the above more realistic cases because at $t_{\rm
  obs}\,\ll\,t_{\rm dec}$, where $t_{\rm dec}$ marks the onset of the
decelerating \cite{1976PhFl...19.1130B} solution, the overall
non-thermal luminosity is suppressed by some power of $t_{\rm
  obs}/t_{\rm dec}$.

Finally, we discuss briefly the expected degree of polarization.
Theoretical and numerical developments both suggest that the
micro-turbulence in the precursor is two-dimensional isotropic,
confined in the plane transverse to the shock normal
\citep[e.g.][]{1999ApJ...526..697M,2008ApJ...673L..39S}. In the
precursor frame, where  this turbulence is mostly of magnetic nature,
the electron distribution is sharply beamed into an opening angle
$1/\Gamma_{\rm sh}$. In such a configuration, the fraction of linearly
polarized synchrotron radiation can be expressed as
\citep{1980MNRAS.193..439L} $p_{\rm lin}\,=\,p_{\rm
  lin,max}f\left(\theta_{\boldsymbol{k}\boldsymbol{\beta_{\rm
      sh}}}\right)$, where $p_{\rm lin,max}$ is the degree of maximum
linear polarization (in a regular field) and
$\theta_{\boldsymbol{k}\boldsymbol{\beta_{\rm sh}}}$ represents the
angle between the direction of propagation of the photon
$\boldsymbol{k}$ and the shock normal $\boldsymbol{\beta_{\rm
    sh}}$. The function $f$ depends on the spectral index of the
electron distribution; the result
$f(\theta)\,=\,\sin^2\theta/\left(1+\cos^2\theta\right)$,
corresponding to $s=3$, is often used as an approximation. However, in
our case, $\theta_{\boldsymbol{k}\boldsymbol{\beta_{\rm
      sh}}}\,\lesssim\,1/\Gamma_{\rm sh}$ implies that the fraction of
linear polarization is at most of order $1/\Gamma_{\rm sh}^2$. This
stands in contrast with the contribution from the downstream plasma,
in which partial linear polarization may subsist
\citep[e.g.][]{2016MNRAS.455.1594N}. One can check that the degree of
circular polarization scales in proportion to $\cos\alpha_{\delta B}$,
where $\alpha_{\delta B}$ represents the angle of the (random)
magnetic field in the transverse plane at a given point; consequently,
this contribution vanishes in the average over $\alpha_{\delta
  B}$. The synchrotron radiation in a Weibel-like micro-turbulent
precursor is thus essentially unpolarized.

In contrast, linear polarization is maximum in the background magnetic
field, with $p_{\rm
  lin,max}\,=\,[7+3(s-1)]/[3+3(s-1)]\,\simeq\,63\,$\% for $s=2.3$ and
regular deflection. This fraction is now much larger than that
expected from the downstream side, even if the field were regular
there, because of the averaging procedures over the viewing angles
that results from the Lorentz boosting of the downstream emission, as
discussed in detail in \citet{2016MNRAS.455.1594N}. In principle, this
offers a new way to study the precursor at the highest frequencies,
where model $(1)$ is expected to hold; unfortunately, the strong
suppression of precursor emission in this model in all cases
considered renders the detection of this polarized signal unlikely.

We note that the beaming of the electron distribution also produces net
circular polarization in the background magnetic field. The fraction
of circularly polarized synchrotron light scales as $p_{\rm
  circ}\,\sim\,1/\langle\gamma\rangle\,{\rm d}\ln \varphi/{\rm
  d}\theta$, where $\varphi$ characterizes the angular part of the
distribution function, $\theta$ represents the pitch angle relative to
the magnetic field, and $\langle\gamma\rangle$ the mean Lorentz
factor. In the precursor, ${\rm d}\ln \varphi/{\rm
  d}\theta\,\sim\,\mathcal{O}\left(\Gamma_{\rm sh}\right)$, but
$\langle\gamma\rangle\,\sim\,\Gamma_{\rm sh}\gamma_{\rm m}$ implies
that the overall fraction remains quite small, of order $1/\gamma_{\rm
  m}$. Here as well, this contribution appears negligible; it could
not, in particular, explain the amount of circular polarization
inferred recently in a gamma-ray burst
afterglow~\citep{2014Natur.509..201W}.

\section{Synchrotron precursor of non-relativistic collisionless shock waves}\label{sec:nrel}
\begin{figure}[ht]
\centerline{\includegraphics[width=0.6\columnwidth]{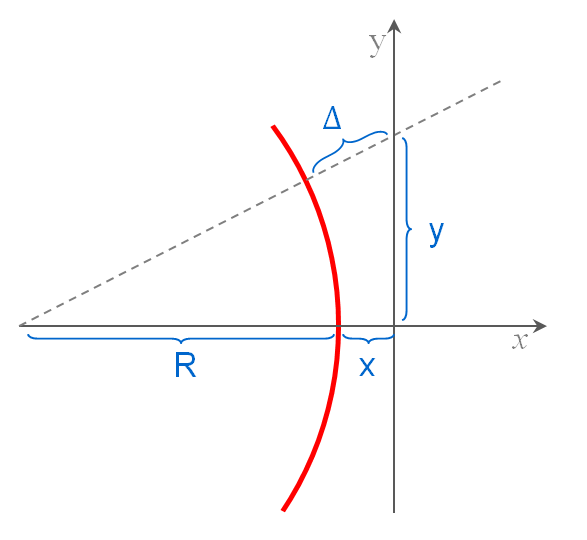}}
  \caption{Sketch of the geometry for studying and computing the
    emission of the precursor in a non-relativistic, supernova
    remnant-like source. The line of sight is oriented along the
    $\boldsymbol{y}-$axis.
    \label{fig:sk-nrel}}
\end{figure}

The sub-relativistic flow velocities seen for example in supernovae
remnants, imply no Lorentz boosting/beaming effects. This is a crucial
distinction because it implies that the source can be seen off-axis,
and in particular that the shock can be imaged edge-on, at its
limb. Such morphological data offers the possibility of isolating the
precursor contribution by concentrating on the emission in front of
the shock, as sketched in Fig.~\ref{fig:sk-nrel}. A major question is
whether one actually sees the precursor at the limb, or some
contribution from shocked plasma further behind due to some departures
from spherical symmetry~\citep{2013ASPC..470..269V}. We leave this
question aside here and compute the morphological and spectral
properties of this upstream synchrotron emission assuming ideal
spherical symmetry. We also neglect the contribution of bremsstrahlung
emission in the X-ray domain, the domain that we are interested in
here.

More specifically, we compute the surface brightness of the
synchrotron radiation
\begin{equation}
  S_{\nu,\rm p}\,=\,\int_{-\infty}^{+\infty}{\rm d}y\,j_\nu
  \label{eq:snu1}
\end{equation}
with $j_\nu$ the position dependent emissivity in the precursor. The
integration is taken along the line of sight, oriented along the
$y-$axis (see  Fig.~\ref{fig:sk-nrel}). We consider a point, located a
distance $x$ in front of the shock, with $x\,\ll\,R$ ($R$ the shock
radius). Then, the coordinate along the line of sight can be written
$y\,=\,\left[\left(R+\Delta \right)^2 -
  \left(R+x\right)^2\right]^{1/2}\,\simeq\,\left[2(\Delta-x)
  R\right]^{1/2}$ in terms of the depth $\Delta$ in the precursor,
measured relative to the shock front radius, as indicated in
Fig.~\ref{fig:sk-nrel}. The integral in Eq.~(\ref{eq:snu1}) thus
samples depths in the precursor $\Delta\in [x,+\infty [$, with
    differing contributions to the emissivity at a given frequency.

We model the distribution function of electrons in the precursor as 
\begin{equation}
\frac{{\rm d}n_e(\Delta)}{{\rm d}p}\,\simeq\,\frac{n_e}{p_{\rm
    min}}\left(\frac{p}{p_{\rm min}}\right)^{-s}\exp\left(-\frac{p}{p_{\rm
    max}}\right)\exp\left[-\frac{\Delta}{r_{\rm p}(p)}\right]
  \label{eq:dndp}
\end{equation}
in terms of the minimum ($p_{\rm min}$) and maximum ($p_{\rm max}$)
momenta of the distribution.  As before, $r_{\rm p}(p)$ represents the
typical length scale of the precursor for particles of momentum
$p$. While the dependence $\propto\,\exp(-x/r_{\rm p})$ represents the
generic solution of the diffusion--advection equation in a uniform
turbulence (see e.g. ~\citealt{1983RPPh...46..973D}), the dependence
$\propto\,\exp(-p/p_{\rm max})$ is phenomenological. The results
derived below can be extended to other cut-off forms discussed in
\cite{2007A&A...465..695Z}.

The precursor length scale is defined by $r_{\rm p}\,\simeq\, ct_{\rm
  scatt}/\beta_{\rm sh}$ in terms of the scattering time and shock
velocity. For a turbulence of homogeneous strength and Bohm scaling of
the scattering frequency, this length scale is $r_{\rm
  p}(p)\,\simeq\,r_{\rm g}/\beta_{\rm sh}$, with $r_{\rm
  g}\,=\,pc/(eB_{\rm p})$ the gyroradius and $B_{\rm p}$ the magnetic
field strength in the precursor; for Kolmogorov-type (homogeneous)
turbulence, however, $r_{\rm p}\,\simeq\,\lambda_{\rm max}^{2/3}r_{\rm
  g}^{1/3}/\beta_{\rm sh}$~\citep[e.g.][]{2002PhRvD..65b3002C}, with
$\lambda_{\rm max}$ the length scale of the turbulence. In order to
allow a possible deviation from a Bohm scaling, we adopt a
phenomenological scaling $r_{\rm
  p}\,\propto\,p^{1/(1+\alpha_\lambda)}$
(i.e. $\alpha_\lambda\,=\,(0,2)$ for  homogeneous Bohm or
Kolmogorov-type scalings, respectively). It may be useful to recall that in a
generic turbulence of scale $\lambda_{\rm max}$, particles of
gyroradius $r_{\rm g}\,\sim\,\lambda_{\rm max}$ indeed scatter at a
rate close to Bohm, but particles of much smaller gyroradius may
scatter at a quite different rate.

It might also be possible to take into account a spatial dependence of
the magnetic field strength in the precursor. However, modelling of
supernova remnants suggests that the pre-shock amplification factor of
the background magnetic field is of the order of a factor of ten
~\citep[e.g.][]{2003A&A...412L..11B}. Therefore, such dependence, if
it exists, is likely to be quite mild. In the following, we assume
that on the scales probed by the electrons, the magnetic field
strength is uniform in the precursor, but we allow for a possible
additional amplification of the latter in the shock transition.

We write the synchrotron spectral power of one electron as
$r_\nu\exp(-\nu/\nu_{\rm c})\sqrt{3}e^3B_{\rm p}/(m_e c^2)$ with
$\nu_{\rm c}\,\propto\,p^2$ the synchrotron peak frequency; the
function $r_\nu\,\simeq\,(\nu/\nu_{\rm c})^{1/3}$ at
$\nu\,\ll\,\nu_{\rm c}$ and $r_\nu\,\sim\,1$ at
$\nu\,\gtrsim\,\nu_{\rm c}$, see Eq.~(30) of
\cite{2007A&A...465..695Z}.

Noting that
\begin{equation}
  \int_{-\infty}^{+\infty}{\rm d}y\,\exp\left(-\Delta/r_{\rm
    p}\right)\,=\,\sqrt{2\pi R r_{\rm p}}\, \exp\left(-x/r_{\rm
    p}\right),
\end{equation}
the surface brightness at position $x\,>\,0$ can be written
\begin{eqnarray}
  S_{\nu}&\,\simeq\,&\frac{1}{4\pi}\sqrt{2\pi R r_{\rm p,min}}
  \frac{\sqrt{3}n_ee^3B_{\rm p}}{m_e c^2}\nonumber\\
  &&\times\int_1^{+\infty}{\rm d}\hat p\,\hat p^{-s + (1+\alpha_\lambda)^{-1}/2}\,r_\nu\,\exp\left(-\frac{x}{r_{\rm p}}-\frac{\nu}{\nu_{\rm c}}-\hat p\frac{p_{\rm min}}{p_{\rm max}}\right)\nonumber\\
  &&
  \label{eq:snu1}
\end{eqnarray}
with $\hat p\,=\, p/p_{\rm min}$, $r_{\rm p,min}\,\equiv\,r_{\rm
  p}(p_{\rm min})$. The exponent $(1+\alpha_\lambda)^{-1}/2$ arises
through the effective depth of integration on the line of sight and
its momentum dependence. In the following, we use the short-hand
notation: $s_0\,\equiv\,\left[R r_{\rm
  p,min}/(2\pi)\right]^{1/2}\sqrt{3}n_ee^3B_{\rm p}/(m_ec^2)$.

The integral in Eq.~(\ref{eq:snu1}) can be integrated in various
limits. Consider first the case $x\,\ll\,r_{\rm p}(p_\nu)$ and
$\nu\,\ll\,\nu_{\rm max}$ with $\nu_{\rm max}\,\equiv\,\nu_{\rm
  c}(p_{\rm max})$. Here, $p_\nu$ is understood as the momentum for
which the synchrotron peak frequency equals $\nu$ and it should not be
confused with the spectral power of an electron, i.e. $\nu_{\rm
  c}(p_\nu)\,\equiv\,\nu$. The above limit then means that at location
$x$,   electrons radiate at their peak frequency at $\nu$;
consequently, the spectrum at $\nu$ will be shaped by these electrons.
The limit $x\,\ll\,r_{\rm p}(p_\nu)$ implies $\exp\left(-x/r_{\rm
  p}\right)\,\sim\,1$ in the range of momenta which contribute to the
radiation at $\nu$, hence this exponential dependence can be safely
ignored. This results in  a usual integration of the synchrotron
spectral power over a powerlaw distribution, albeit with a modified
index $s-(1+\alpha_\lambda)^{-1}/2$, which leads to the result
(omitting numerical prefactors of the order of unity)
\begin{equation}
  S_\nu\,\simeq\,s_0 \left(\frac{\nu}{\nu_{\rm min}}\right)^{(1-s)/2 +
    (1+\alpha_\lambda)^{-1}/4} \quad\quad \left[x\,\ll\,r_{\rm p}(p_\nu)\right]
  \label{eq:snup1}
\end{equation}
We note that the constraint $x\,\ll\,r_{\rm p}(p_\nu)$ can also be
rewritten $\nu\,\gg\,\nu_<(x)$ with $\nu_<(x)$ defined implicitly by
$r_{\rm p}\left[p_{\nu_<(x)}\right]\,=\,x$. Its meaning is the
following: at distance $x$, the electron distribution has an effective
low-energy cut-off at $p$ such that $r_{\rm p}(p)\,=\,x$ because
particles of smaller momenta cover smaller distances in the precursor;
hence, the local minimum synchrotron frequency of this population is
$\nu_<(x)$.

We consider now the limit $x\,\gg\,r_{\rm p}(p_\nu)$, with still
$\nu\,\ll\,\nu_{\rm max}$. In this limit, $\nu\,\ll\,\nu_<(x)$, hence
at all $\Delta\,>\,x$, we collect at $\nu$ only the low-frequency tail
$\,\propto\,\nu^{1/3}$ of the spectral flux of high-energy particles.
To model the flux in this limit, it is possible to make the
approximation $\exp\left(-x/r_{\rm p}\right)\,\sim\,\Theta\left(r_{\rm
  p}-x\right)$ as in Sec.~\ref{sec:rel}. This Heaviside function in
$x$ effectively constrains the range of integration, i.e.
Eq.~(\ref{eq:snu1}) can be rewritten
\begin{eqnarray}
  S_{\nu}&\,\simeq\,&s_0\,
  \int_{(x/r_{\rm p,min})^{(1+\alpha_\lambda)}}^{+\infty}{\rm d}\hat p\,\hat p^{-s + (1+\alpha_\lambda)^{-1}/2}\,\left(\frac{\nu}{\nu_{\rm c}}\right)^{1/3}\nonumber\\
  &\,\simeq\,& s_0\,\left(\frac{x}{r_{\rm p,min}}\right)^{\left(\frac{1}{3}-s\right)(1+\alpha_\lambda)+\frac{1}{2}}\left(\frac{\nu}{\nu_{\rm min}}\right)^{1/3}\quad\left[x\,\gg\,r_{\rm p}(p_\nu)\right]\nonumber\\
  &&
  \label{eq:snup2}
\end{eqnarray}
To summarize, the general spectro-morphological aspect of the
synchrotron surface brightness of the precursor well below the maximum 
frequency $\nu_{\rm max}$ is given by the conjunction of
Eqs.~(\ref{eq:snup1}) and (\ref{eq:snup2}). At $x\,\ll\,r_{\rm
  p}(p_\nu)$, the surface brightness is constant in $x$ with a harder
than usual spectrum in $\nu$, then decreases according to
Eq.~(\ref{eq:snup2}) at $x\,\gg\,r_{\rm p}(p_\nu)$,
e.g. $\,\propto\,x^{-7/6}$ for $s=2$ and $\alpha_\lambda\,=\,0$.

As we will see, the detection of the precursor becomes promising only
at the highest frequencies, which, for typical supernovae remnants,
fall in the X-ray range. Indeed, electrons radiating at $\nu$ at their
synchrotron peak frequency have an energy
$\epsilon_e\,\simeq\,10\,{\rm TeV}\, (h\nu/1\,{\rm
  keV})^{1/2}n_0^{-1/4}\beta_{\rm sh, -1.5}^{-1/2}\epsilon_{B\rm
  p,-3}^{-1/4}$, where $\epsilon_{B\rm p}\,\equiv\,B_{\rm
  p}^2/\left(4\pi n m \beta_{\rm sh}^2c^2\right)$. As  for the maximum
energy of accelerated electrons, balancing synchrotron cooling with
the acceleration time scale $t_{\rm acc}\,=\,\kappa\,r_{\rm
  p}/(\beta_{\rm sh} c)$, we find
\begin{eqnarray}
  \left.\epsilon_{e,\rm max}\right\vert_{\alpha_\lambda=0}
  &\,\simeq\,&50\,{\rm TeV}\,\,\beta_{\rm sh,-1.5}^{1/2}n_0^{-1/4}\epsilon_{B\rm p,-3}^{-1/4}\kappa_1^{-1/2}\nonumber\\
\left.\epsilon_{e,\rm max}\right\vert_{\alpha_\lambda=1}
  &\,\simeq\,&25\,{\rm TeV}\,\,\beta_{\rm sh,-1.5}^{1/3}n_0^{-1/2}\epsilon_{B\rm p,-3}^{-1/2}\kappa_1^{-2/3}\lambda_{\rm max,16}^{-1/3}
\end{eqnarray}

It is thus necessary  to compute the spectrum close to $\nu_{\rm max}$ and at
distances $x\,\sim\,r_{\rm p,max}$ $\left[r_{\rm
    p,max}\,\equiv\,r_{\rm p}(p_{\rm max})\right]$. In this limit,  the exponential turn-over of the electron distribution in
momentum and in distance must be retained as the exponential turn-over of the
synchrotron spectral power. It is possible nevertheless obtain a reasonable
approximation to $S_\nu$ through a saddle-point approximation of the
integral. To this effect, we define
\begin{equation}
  g(p)\,=\,\frac{x}{r_{\rm p}}+\frac{\nu}{\nu_{\rm c}}+\frac{p}{p_{\rm max}}
\end{equation}
and $p_\star$ such that $g'(p_\star)\,=\,0$. Then
\begin{eqnarray}
  S_\nu&\,\simeq\,&s_0\left(\frac{p_\star}{p_{\rm min}}\right)^{1-s+(1+\alpha_\lambda)^{-1}/2}\sqrt{\frac{2\pi}{g''_\star}}r_\nu(p_\star)\,e^{-g_\star}\nonumber\\
  && \quad\quad\left[x\,\gtrsim\,r_{\rm p,max}\,\lor\,\nu\,\gtrsim\,\nu_{\rm max}\right]
  \label{eq:snup3}
\end{eqnarray}
with the short-hand notations: $g_\star\,\equiv\,g(p_\star)$,
$g''_\star\,=\,g''(p_\star)$. In the limits $x\,\gtrsim\,r_{\rm
  p,max}\,\land\,\nu\,\ll\,\nu_{\rm max}$, or $x\,\ll\,r_{\rm
  p,max}\,\land\,\nu\,\gtrsim\,\nu_{\rm max}$,  the
following approximate solutions can be used:
\begin{eqnarray}
  p_\star&\,\simeq\,&p_{\rm max}\,\,{\rm max}\left[1.2\left(\frac{\nu}{\nu_{\rm
        max}}\right)^{1/3},\,\,a_\lambda\left(\frac{x}{r_{\rm
        p,max}}\right)^{(1+\alpha_\lambda)/(2+\alpha_\lambda)}\right]\nonumber\\
  g_\star&\,\simeq\,&{\rm
    max}\left[1.9\left(\frac{\nu}{\nu_{\rm
        max}}\right)^{1/3},\,\,b_\lambda\left(\frac{x}{r_{\rm
        p,max}}\right)^{(1+\alpha_\lambda)/(2+\alpha_\lambda)}\right]\nonumber\\
  g''_\star&\,\simeq\,&{\rm
    min}\left[2.4\left(\frac{\nu}{\nu_{\rm
        max}}\right)^{-1/3},\,\,c_\lambda\left(\frac{x}{r_{\rm
        p,max}}\right)^{-(1+\alpha_\lambda)/(2+\alpha_\lambda)}\right]
  \label{eq:snuhi2}
\end{eqnarray}
with
$a_\lambda\,=\,(1+\alpha_\lambda)^{-(1+\alpha_\lambda)/(2+\alpha_\lambda)}$,
$b_\lambda\,=\,(1+\alpha_\lambda)^{1/(2+\alpha_\lambda)} + a_\lambda$,
and
$c_\lambda\,=\,(2+\alpha_\lambda)(1+\alpha_\lambda)^{-1/(2+\alpha_\lambda)}$.
With these approximate scalings, Eq.~(\ref{eq:snup3}) provides the
leading order approximation to the spectro-morphological shape of the
precursor emission at high frequency and/or great distance. It
indicates in particular that the precursor emission can extend up to a
few times $r_{\rm p,max}$ with a rather smooth exponential decay
beyond.

The contribution from the shocked plasma exists only at negative
values of $x$.  At a given $x$,   the (homogeneous)
contribution of this shocked plasma is integrated up to an effective depth along the
line of sight $y\,\simeq\,\sqrt{2R\vert x\vert}$ for $\vert
x\vert\,\ll\,R$. Repeating the above integrations, the downstream
surface brightness can thus be written as
\begin{equation}
  S_{\nu,\rm d}\,\simeq\, r_{\rm amp}^{(s+1)/2}\,s_0\,\left(\frac{\vert
    x\vert}{x_{\rm min}}\right)^{1/2}\, \left[\frac{\nu}{\nu_{\rm
        min}}\right]^{(1-s)/2} \quad\left[\nu\,\ll\,r_{\rm amp}\nu_{\rm max}\right]
\label{eq:snud1}
\end{equation}
The first term on the right-hand side, with $r_{\rm amp}\,\simeq\,4A_{\rm amp}$,
contains both the compression ratio at the shock ($4$) and a possible
amplification factor due to the non-linear processing downstream of
the shock ($A_{\rm amp}$). The corresponding exponent $(s+1)/2$ arises
because the spectrum has been expressed in terms of $\nu/\nu_{\rm
  min}$, with $\nu_{\rm min}$ the minimum frequency in the pre-shock
magnetic field. The above expression should  cut off beyond
the cooling length $l_{\rm cool}$ of particles radiating at frequency
$\nu$. As discussed in \cite{2004A&A...419L..27B}, $l_{\rm
  cool}\,\simeq\,{\rm max}\left[\sqrt{l_{\rm d}l_{\rm c}},l_{\rm
    c}\right]$ in terms of the downstream diffusive length scale
$l_{\rm d}\,\simeq\,l_{\rm scatt}/\beta_{\rm d}$ ($\beta_{\rm
  d}\,=\,\beta_{\rm sh}/4$ the downstream shock velocity), $l_{\rm
  scatt}$ the mean free path, and $l_{\rm c}\,=\,\beta_{\rm d}c t_{\rm
  cool}$ the convective cooling length scale.

In the turn-over regime, $\nu\,\gtrsim\,r_{\rm amp}\,\nu_{\rm max}$ is
 derived
\begin{eqnarray}
 S_{\nu,\rm d}&\,\simeq\,& r_{\rm amp}\,s_0\,\left(\frac{\vert
   x\vert}{x_{\rm min}}\right)^{1/2}\, \left(\frac{p_{\star,\rm
     d}}{p_{\rm min}}\right)^{1-s}\sqrt{\frac{2\pi}{g''_{\star,\rm
       d}}}r_\nu(p_{\star,\rm d})\,e^{-g_{\star,\rm d}}\nonumber\\
 &&  \quad\quad\left[\nu\,\gtrsim\,r_{\rm amp}\nu_{\rm max}\right]
\label{eq:snud2}
\end{eqnarray}
and $p_{\star,\rm d}$, $g_{\star,\rm d}$, and $g''_{\star,\rm d}$ can
be obtained in the same way as their upstream counterparts in
Eq.~(\ref{eq:snuhi2}), with the replacements $x\,\rightarrow\,0$,
$\alpha_\lambda\,\rightarrow\,0$, and $\nu_{\rm
  max}\,\rightarrow\,r_{\rm amp}\nu_{\rm max}$.

\begin{figure}[ht]
\centerline{\includegraphics[width=0.9\columnwidth]{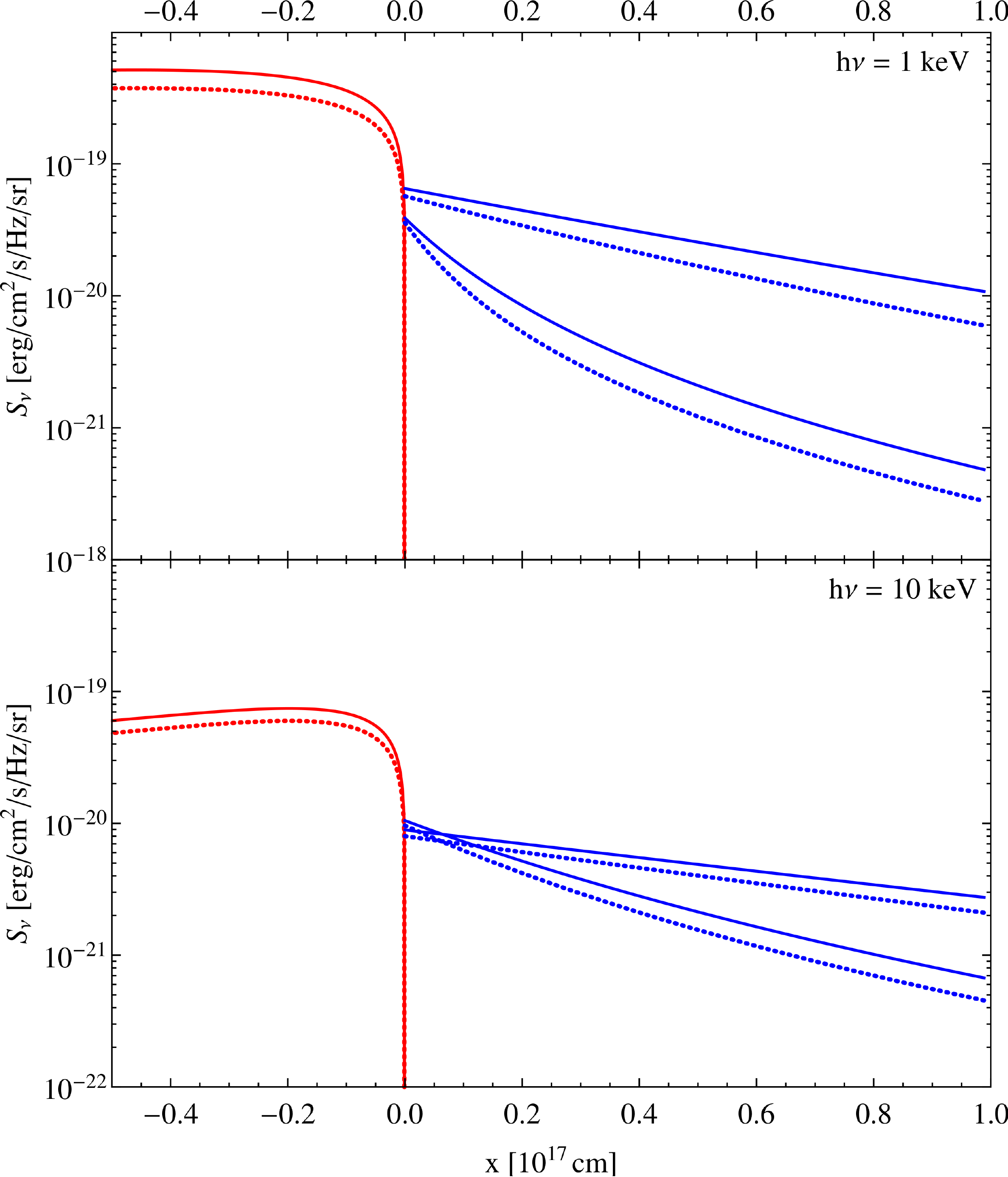}}
  \caption{Surface brightness along the line of sight to the limb of a
    non-relativistic shock as a function of distance $x$ to the
    shock, for various cases as shown: upper panel at
    $h\nu\,=\,1\,$keV, lower panel at $h\nu=10\,$keV. The solid lines
    present a complete numerical evaluation of Eq.~(\ref{eq:snu1}),
    while the dotted lines correspond to the analytical approximation
    in the turn-over regime. Two models are shown for the precursor
    emission: $\alpha_\lambda\,=\,1$ for the upper lines, i.e. $t_{\rm
      scatt}\,\propto\,p^{1/2}$, and $\alpha_\lambda\,=\,0$ for the
    lower lines, i.e. Bohm scaling $t_{\rm scatt}\,\propto\,p$. In all
    cases, $\beta_{\rm sh}\,=\,1/30$, $n_{\rm u}\,=\,1\,$cm$^{-3}$,
    $R\,=\,1\,$pc, $\epsilon_{B{\rm p}}\,=\,0.001$, and
    $\epsilon_e\,=\,0.01$.
    \label{fig:prec-nrel}}
\end{figure}

Figure~\ref{fig:prec-nrel} presents examples of the morphological
dependence of the precursor which illustrate the above scalings. In
these figures, the synchrotron cooling of high-energy particles on the
downstream side has been explicitly taken into account using the
spatial profile described in \citet{2004A&A...419L..27B}.

Figure~\ref{fig:prec-nrel} indicates that the contribution from the
precursor is sizeable at the highest frequencies, especially if
maximum energy electrons scatter more slowly than the Bohm scaling. To
put things in proper perspective, recall that $1''$ resolution at a
distance of $1\,$kpc corresponds to a spatial resolution of $1.4\times
10^{16}\,$cm, which provides a reasonable approximation to the extent
of the precursor emission $r_{\rm p,max}$. If the magnetic field were
to receive substantial amplification in the near precursor, i.e. at
distances $x\,\ll\,r_{\rm p,max}$, the synchrotron contribution of the
precursor would be accordingly reduced.

Using Eqs.~(\ref{eq:snup1}), (\ref{eq:snup2}), (\ref{eq:snup3}), 
(\ref{eq:snud1}), and (\ref{eq:snud2}), it is possible to compare the emission of
the precursor relative to that of downstream. In the inertial range
$\nu\,\ll\,\nu_{\rm max}$, the ratio of the upstream contribution, at
its maximum, relative to the downstream contribution, is of order
$r_{\rm amp}^{-(s+1)/2}\left[r_{\rm p}(p_\nu)/l_{\rm
    cool}\right]^{1/2}$.  In the turn-over regime,
$\nu\,\gtrsim\,\nu_{\rm max}$, this ratio becomes of the order of
$r_{\rm amp}^{-1}\left[r_{\rm p,max}/l_{\rm cool}\right]^{1/2}$ up to
a factor of the order of unity which is related to the difference of
$\exp(-g_\star)$ between upstream and downstream.  

For reference,
\begin{eqnarray}
  l_{\rm c}&\,\simeq\,&
  6\times10^{16}n_0^{-3/4}A_{\rm amp}^{-5/2}\beta_{\rm sh,-1.5}^{-1/2}\epsilon_{B{\rm
      p},-3}^{-3/4}\left(\frac{h\nu}{1\,{\rm keV}}\right)^{-1/2} \,\,{\rm cm}
  \nonumber\\
  \left.l_{\rm d}\right\vert_{\alpha_\lambda=0}&\,\simeq\,&4\times 10^{15}\,n_0^{-3/4}A_{\rm
    amp}^{-1/2}\beta_{\rm sh,-1.5}^{-5/2}\epsilon_{B{\rm
      p},-3}^{-3/4}\left(\frac{h\nu}{1\,{\rm keV}}\right)^{1/2}\,\,{\rm cm}\nonumber\\
 \left.l_{\rm d}\right\vert_{\alpha_\lambda=1}&\,\simeq\,&6\times 10^{16}\,n_0^{-3/8}A_{\rm
    amp}^{1/4}\beta_{\rm sh,-1.5}^{-7/4}\epsilon_{B{\rm
      p},-3}^{-3/8}\lambda_{\rm max,16}^{1/2}\left(\frac{h\nu}{1\,{\rm keV}}\right)^{1/4}
      {\rm cm}\nonumber\\
      \label{eq:lcool1}
\end{eqnarray}
Furthermore,
\begin{eqnarray}
\left.r_{\rm p}(p_\nu)\right\vert_{\alpha_\lambda=0}&\,\simeq\,&0.7\times 10^{16}\,n_0^{-3/4}\beta_{\rm  sh,-1.5}^{-5/2}\epsilon_{B\rm p,-3}^{-3/4} \left(\frac{h\nu}{1\,{\rm
    keV}}\right)^{1/2}\,\,{\rm cm}\nonumber\\
\left.r_{\rm p}(p_\nu)\right\vert_{\alpha_\lambda=1}&\,\simeq\,& 5\times
10^{16}\,n_0^{-3/8}\beta_{\rm sh,-1.5}^{-7/4}\epsilon_{B\rm p,-3}^{-3/8}\lambda_{\rm
  max,16}^{1/2}\left(\frac{h\nu}{1\,{\rm keV}}\right)^{1/4}{\rm cm}\nonumber\\
\left.r_{\rm p,max}\right\vert_{\alpha_\lambda=0}&\,\simeq\,&3.6\times 10^{16}n_0^{-3/4}\beta_{\rm sh,-1.5}^{-3/2}\epsilon_{B\rm p,-3}^{-3/4}\kappa_1^{-1/2}\,\,{\rm cm}\nonumber\\
\left.r_{\rm p,max}\right\vert_{\alpha_\lambda=1}&\,\simeq\,&
7\times 10^{16}\,n_0^{-1/2}\beta_{\rm sh,-1.5}^{-4/3}\epsilon_{B\rm p,-3}^{-1/2}
\kappa_1^{-1/3}\lambda_{\rm max,16}^{1/3}\,\,{\rm cm}\nonumber\\
&&
\label{eq:dmaxn}
\end{eqnarray}
The ``post-precursor'' amplification factor $A_{\rm amp}$ exerts two
opposite influences on the relative magnitude of the precursor
emission: if $A_{\rm amp}\,>\,1$, the downstream cooling length
decreases ($l_{\rm c}\,\propto\,A_{\rm amp}^{-5/2}$ and $l_{\rm
  d}\,\propto\,A_{\rm amp}^{-1/2}$), which effectively reduces the
depth of integration of the downstream emission on the line of sight,
but in the meantime the downstream emissivity increases by a factor
$A_{\rm amp}^{3/2}$. 

Quite interestingly, the addition of polarimetric data to the above
spectro-morphological characteristics could prove an extremely useful
tool in extracting the contribution of the precursor. One may indeed
expect, on rather general grounds, that the magnetic field in the
precursor would be more amplified in the perpendicular directions than
along the shock normal because the cosmic-ray current along this
shock normal exerts a force perpendicular to it and to the background
magnetic field~\citep{2004MNRAS.353..550B}. Whether the overall
turbulence is indeed anisotropic  in the far precursor, where
most of the precursor synchrotron flux is emitted, is however
currently
uncertain~\citep{2010ApJ...717.1054R,2012MNRAS.419.2433R,2013MNRAS.430.2873R,2013ApJ...765L..20C,2014ApJ...794...46C}. One
may decompose the magnetic turbulence energy density in its
contribution parallel to the shock normal, $\epsilon_{B\parallel}$,
and its contribution perpendicular to the shock normal
$\epsilon_{B\perp}$; a field mostly tangled in the plane transverse to
the shock normal is thus such that
$\eta_B\,\equiv\,2\,\epsilon_{B\parallel}/\epsilon_{B\perp}\,<\,1$. If
this limit holds, then the synchrotron signal will be about maximally
linearly polarized when seen at an angle $\Delta\theta\,\simeq\,\pi/2$
relative to the shock normal, which corresponds to the viewing angle
at the limb of a shock front~\citep{1980MNRAS.193..439L}. In contrast,
one expects the downstream turbulence to be mostly isotropized through
shock crossing, except for localized small-scale polarization
clumps~\citep{2009MNRAS.399.1119B}.

The detection of polarization in the shock vicinity might thus provide
a net signature of the precursor, and for the same reasons, it could
allow the morphological extension of the precursor to be distinguished
from a possible corrugation of the shock front. A detailed study,
accounting for the finite resolution of an experiment, would be needed
to specify to which level the precursor signal could be extracted, but
as an order of magnitude, angular resolutions not far above $1''$
appear necessary, given the above estimates.  Numerical simulations of
the instability in the non-linear regime also appear to be mandatory
to properly characterize the expected degree of polarization.

\section{Summary and conclusions}\label{sec:conc}
Many of the nagging questions concerning the acceleration of particles
at shock waves find their answers in the physics of the precursor of
these shocks where the accelerated particles interact with the
unshocked plasma, thereby seeding the electromagnetic instabilities
that shape the collisionless shock and the acceleration process
itself. The direct or indirect observation of the synchrotron
radiation from accelerated particles in this precursor might therefore
offer a new means to study these physics. This paper has discussed the
possibility of such a detection, in relativistic and non-relativistic
collisionless shocks.

We reach the general conclusion that such a detection is arduous. For
relativistic outflows, one has to extract the synchrotron spectral
contribution of the precursor from the overall synchrotron
self-Compton spectrum of the source because the shock is seen
head-on. Interestingly, we find that this precursor spectrum is
generally quite hard because the upstream residence time increases
with particle energy, so that the spectrum of accelerated particles as
integrated over the whole precursor is itself quite hard. At the
minimum frequency of the synchrotron spectrum, the relative
contribution of the precursor is tiny, of order $\Delta/R$, where
$\Delta$ corresponds to the microphysical length scale explored by
particles of the minimum Lorentz factor, $R$ to the shock radius, and
$\Gamma_{\rm sh}$ to the bulk Lorentz factor of the shock. Detection
could thus be expected only at the highest energies, typically in the
X-ray range. However, we find that in practice, even at these high
energies, the synchrotron contribution of the precursor is dominated
either by the synchrotron or by the inverse Compton spectrum of the
shocked plasma. In some extreme cases, in particular for
trans-relativistic shock waves, one could hope to detect the
contribution of the precursor if the small-scale self-generated
turbulence dominates the transport of particles up to the highest
energies.

In contrast, in non-relativistic sources, the shock front can be
observed edge-on, and therefore the precursor might be extracted from
the morphology of the synchrotron surface brightness. We have
explicitly calculated this spectro-morphological signature at a given
frequency $\nu$, finding that it extends away from the shock up to the
distance $r_{\rm p}$ to which particles radiating their synchrotron
peak frequency at $\nu$ can propagate, and then it decreases beyond
this distance. We also note also the precursor spectrum is harder than
the standard synchrotron in this regime, and that $r_{\rm p}$
increases with frequency. For reference and for parameters typical of
a supernova remnant, this distance $r_{\rm p,max}$ is of the order of
$10^{16}\,$cm at the maximum energies, corresponding to radiation in
the X-ray range. Such a scale represents a separation of $1''$ on the
sky at a distance of $1\,$kpc. Close to the maximum synchrotron photon
energy, the typical ratio of the emission of the precursor relative to
that of the shocked plasma is of order $r_{\rm amp}^{-1}(r_{\rm
  p,max}/l_{\rm cool})^{1/2}$, where $r_{\rm amp}\,\sim\,4\,A_{\rm
  amp}$ is the shock amplification factor of the magnetic field,
$s\,\sim\,2$ the index of the accelerated particle spectrum, and
$l_{\rm cool}$ the cooling length of particles behind the shock
front. The detection of the precursor thus appears challenging, but
not altogether excluded.

An interesting consequence of viewing the shock edge-on is that the
turbulence, if tangled in the plane transverse to the shock normal in
the precursor, is expected to leave a signature of near maximally
linearly polarized synchrotron light. Polarimetric data with an
adequate angular resolution then emerges as an essential tool to
extract the contribution of the precursor and, in particular, to
distinguish it from potential shock corrugation.

\begin{acknowledgements}
This work has been financially supported by the Programme National
Hautes \'Energies (PNHE) of the C.N.R.S. and by the ANR-14-CE33-0019
MACH project.
\end{acknowledgements}

\bibliographystyle{aa}
\bibliography{shock.bib}

\begin{thebibliography}{46}
\expandafter\ifx\csname natexlab\endcsname\relax\def\natexlab#1{#1}\fi

\bibitem[{{Achterberg} {et~al.}(1994){Achterberg}, {Blandford}, \&
  {Reynolds}}]{1994A&A...281..220A}
{Achterberg}, A., {Blandford}, R.~D., \& {Reynolds}, S.~P. 1994, \aap, 281, 220

\bibitem[{{Achterberg} {et~al.}(2001){Achterberg}, {Gallant}, {Kirk}, \&
  {Guthmann}}]{2001MNRAS.328..393A}
{Achterberg}, A., {Gallant}, Y.~A., {Kirk}, J.~G., \& {Guthmann}, A.~W. 2001,
  MNRAS, 328, 393

\bibitem[{{Ammosov} {et~al.}(1994){Ammosov}, {Ksenofontov}, {Nikolaev}, \&
  {Petukhov}}]{1994AstL...20..157Az}
{Ammosov}, A.~E., {Ksenofontov}, L.~T., {Nikolaev}, V.~S., \& {Petukhov}, S.~I.
  1994, Astronomy Letters, 20, 157

\bibitem[{{Bell}(2004)}]{2004MNRAS.353..550B}
{Bell}, A.~R. 2004, Month. Not. Roy. Astron. Soc., 353, 550

\bibitem[{{Berezhko} {et~al.}(2003){Berezhko}, {Ksenofontov}, \&
  {V{\"o}lk}}]{2003A&A...412L..11B}
{Berezhko}, E.~G., {Ksenofontov}, L.~T., \& {V{\"o}lk}, H.~J. 2003, AA, 412,
  L11

\bibitem[{{Berezhko} \& {V{\"o}lk}(2004)}]{2004A&A...419L..27B}
{Berezhko}, E.~G. \& {V{\"o}lk}, H.~J. 2004, \aap, 419, L27

\bibitem[{{Blandford} \& {Eichler}(1987)}]{1987PhR...154....1B}
{Blandford}, R. \& {Eichler}, D. 1987, Phys. Rep., 154, 1

\bibitem[{{Blandford} \& {McKee}(1976)}]{1976PhFl...19.1130B}
{Blandford}, R.~D. \& {McKee}, C.~F. 1976, Physics of Fluids, 19, 1130

\bibitem[{{Bykov} {et~al.}(2012){Bykov}, {Gehrels}, {Krawczynski}, {Lemoine},
  {Pelletier}, \& {Pohl}}]{2012SSRv..173..309B}
{Bykov}, A., {Gehrels}, N., {Krawczynski}, H., {et~al.} 2012, Sp. Sc. Rev.,
  173, 309

\bibitem[{{Bykov} {et~al.}(2009){Bykov}, {Uvarov}, {Bloemen}, {den Herder}, \&
  {Kaastra}}]{2009MNRAS.399.1119B}
{Bykov}, A.~M., {Uvarov}, Y.~A., {Bloemen}, J.~B.~G.~M., {den Herder}, J.~W.,
  \& {Kaastra}, J.~S. 2009, \mnras, 399, 1119

\bibitem[{{Caprioli} \& {Spitkovsky}(2013)}]{2013ApJ...765L..20C}
{Caprioli}, D. \& {Spitkovsky}, A. 2013, \apjl, 765, L20

\bibitem[{{Caprioli} \& {Spitkovsky}(2014)}]{2014ApJ...794...46C}
{Caprioli}, D. \& {Spitkovsky}, A. 2014, \apj, 794, 46

\bibitem[{{Casse} {et~al.}(2002){Casse}, {Lemoine}, \&
  {Pelletier}}]{2002PhRvD..65b3002C}
{Casse}, F., {Lemoine}, M., \& {Pelletier}, G. 2002, \prd, 65, 023002

\bibitem[{{Chang} {et~al.}(2008){Chang}, {Spitkovsky}, \&
  {Arons}}]{2008ApJ...674..378C}
{Chang}, P., {Spitkovsky}, A., \& {Arons}, J. 2008, ApJ, 674, 378

\bibitem[{{Drury}(1983)}]{1983RPPh...46..973D}
{Drury}, L.~O. 1983, Reports on Progress in Physics, 46, 973

\bibitem[{{Ellison} \& {Cassam-Chena{\"i}}(2005)}]{2005ApJ...632..920E}
{Ellison}, D.~C. \& {Cassam-Chena{\"i}}, G. 2005, \apj, 632, 920

\bibitem[{{Gedalin} {et~al.}(2008){Gedalin}, {Balikhin}, \&
  {Eichler}}]{2008PhRvE..77b6403G}
{Gedalin}, M., {Balikhin}, M.~A., \& {Eichler}, D. 2008, \pre, 77, 026403

\bibitem[{{Ginzburg} \& {Syrovatskii}(1965)}]{1965ARA&A...3..297G}
{Ginzburg}, V.~L. \& {Syrovatskii}, S.~I. 1965, \araa, 3, 297

\bibitem[{{Kelner} {et~al.}(2013){Kelner}, {Aharonian}, \&
  {Khangulyan}}]{2013ApJ...774...61K}
{Kelner}, S.~R., {Aharonian}, F.~A., \& {Khangulyan}, D. 2013, Astrophys. J.,
  774, 61

\bibitem[{{Laing}(1980)}]{1980MNRAS.193..439L}
{Laing}, R.~A. 1980, \mnras, 193, 439

\bibitem[{{Legg} \& {Westfold}(1968)}]{1968ApJ...154..499L}
{Legg}, M.~P.~C. \& {Westfold}, K.~C. 1968, \apj, 154, 499

\bibitem[{{Lemoine}(2013)}]{2013MNRAS.428..845L}
{Lemoine}, M. 2013, MNRAS, 428, 845

\bibitem[{{Lemoine}(2015{\natexlab{a}})}]{2015JPlPh..81a4501L}
{Lemoine}, M. 2015{\natexlab{a}}, Journal of Plasma Physics, 81, 455810101

\bibitem[{{Lemoine}(2015{\natexlab{b}})}]{2015MNRAS.453.3772L}
{Lemoine}, M. 2015{\natexlab{b}}, MNRAS, 453, 3772

\bibitem[{{Lemoine} {et~al.}(2013){Lemoine}, {Li}, \&
  {Wang}}]{2013MNRAS.435.3009L}
{Lemoine}, M., {Li}, Z., \& {Wang}, X.-Y. 2013, Month. Not. Roy. Astron. Soc.,
  435, 3009

\bibitem[{{Li} \& {Waxman}(2006)}]{2006ApJ...651..328L}
{Li}, Z. \& {Waxman}, E. 2006, ApJ, 651, 328

\bibitem[{{Long} {et~al.}(2003){Long}, {Reynolds}, {Raymond}, {Winkler},
  {Dyer}, \& {Petre}}]{2003ApJ...586.1162L}
{Long}, K.~S., {Reynolds}, S.~P., {Raymond}, J.~C., {et~al.} 2003, \apj, 586,
  1162

\bibitem[{{Medvedev} \& {Loeb}(1999)}]{1999ApJ...526..697M}
{Medvedev}, M.~V. \& {Loeb}, A. 1999, ApJ, 526, 697

\bibitem[{{Milosavljevi{\'c}} \& {Nakar}(2006)}]{2006ApJ...651..979M}
{Milosavljevi{\'c}}, M. \& {Nakar}, E. 2006, \apj, 651, 979

\bibitem[{{Morlino} {et~al.}(2010){Morlino}, {Amato}, {Blasi}, \&
  {Caprioli}}]{2010MNRAS.405L..21M}
{Morlino}, G., {Amato}, E., {Blasi}, P., \& {Caprioli}, D. 2010, \mnras, 405,
  L21

\bibitem[{{Nava} {et~al.}(2016){Nava}, {Nakar}, \&
  {Piran}}]{2016MNRAS.455.1594N}
{Nava}, L., {Nakar}, E., \& {Piran}, T. 2016, \mnras, 455, 1594

\bibitem[{{Panaitescu} \& {Kumar}(2000)}]{2000ApJ...543...66P}
{Panaitescu}, A. \& {Kumar}, P. 2000, \apj, 543, 66

\bibitem[{{Pelletier} {et~al.}(2016){Pelletier}, {Bykov}, {Ellison}, \&
  {Lemoine}}]{Pea2016}
{Pelletier}, G., {Bykov}, A., {Ellison}, D., \& {Lemoine}, M. 2016, in prep.

\bibitem[{{Pelletier} {et~al.}(2009){Pelletier}, {Lemoine}, \&
  {Marcowith}}]{2009MNRAS.393..587P}
{Pelletier}, G., {Lemoine}, M., \& {Marcowith}, A. 2009, MNRAS, 393, 587

\bibitem[{{Piran}(2004)}]{2004RvMP...76.1143P}
{Piran}, T. 2004, Rev. Mod. Phys., 76, 1143

\bibitem[{{Reville} \& {Bell}(2012)}]{2012MNRAS.419.2433R}
{Reville}, B. \& {Bell}, A.~R. 2012, MNRAS, 419, 2433

\bibitem[{{Reville} \& {Bell}(2013)}]{2013MNRAS.430.2873R}
{Reville}, B. \& {Bell}, A.~R. 2013, MNRAS, 430, 2873

\bibitem[{{Reynolds}(1998)}]{1998ApJ...493..375R}
{Reynolds}, S.~P. 1998, \apj, 493, 375

\bibitem[{{Riquelme} \& {Spitkovsky}(2010)}]{2010ApJ...717.1054R}
{Riquelme}, M.~A. \& {Spitkovsky}, A. 2010, \apj, 717, 1054

\bibitem[{{Sari} \& {Esin}(2001)}]{2001ApJ...548..787S}
{Sari}, R. \& {Esin}, A.~A. 2001, \apj, 548, 787

\bibitem[{{Sironi} \& {Spitkovsky}(2009)}]{2009ApJ...707L..92S}
{Sironi}, L. \& {Spitkovsky}, A. 2009, Astrophys. J. Lett., 707, L92

\bibitem[{{Spitkovsky}(2008)}]{2008ApJ...673L..39S}
{Spitkovsky}, A. 2008, ApJL, 673, L39

\bibitem[{{Vink}(2013)}]{2013ASPC..470..269V}
{Vink}, J. 2013, in Astronomical Society of the Pacific Conference Series, Vol.
  470, 370 Years of Astronomy in Utrecht, ed. G.~{Pugliese}, A.~{de Koter}, \&
  M.~{Wijburg}, 269

\bibitem[{{Wang} {et~al.}(2010){Wang}, {He}, {Li}, {Wu}, \&
  {Dai}}]{2010ApJ...712.1232W}
{Wang}, X.-Y., {He}, H.-N., {Li}, Z., {Wu}, X.-F., \& {Dai}, Z.-G. 2010, \apj,
  712, 1232

\bibitem[{{Wiersema} {et~al.}(2014){Wiersema}, {Covino}, {Toma}, {van der
  Horst}, {Varela}, {Min}, {Greiner}, {Starling}, {Tanvir}, {Wijers},
  {Campana}, {Curran}, {Fan}, {Fynbo}, {Gorosabel}, {Gomboc}, {G{\"o}tz},
  {Hjorth}, {Jin}, {Kobayashi}, {Kouveliotou}, {Mundell}, {O'Brien}, {Pian},
  {Rowlinson}, {Russell}, {Salvaterra}, {di Serego Alighieri}, {Tagliaferri},
  {Vergani}, {Elliott}, {Fari{\~n}a}, {Hartoog}, {Karjalainen}, {Klose},
  {Knust}, {Levan}, {Schady}, {Sudilovsky}, \&
  {Willingale}}]{2014Natur.509..201W}
{Wiersema}, K., {Covino}, S., {Toma}, K., {et~al.} 2014, \nat, 509, 201

\bibitem[{{Zirakashvili} \& {Aharonian}(2007)}]{2007A&A...465..695Z}
{Zirakashvili}, V.~N. \& {Aharonian}, F. 2007, \aap, 465, 695

\end{thebibliography}

\end{document}